\documentclass[conference,a4paper]{APSIPA2026}
\usepackage{amsmath}
\usepackage{graphicx}
\usepackage{multirow}
\usepackage{threeparttable}
\usepackage{xcolor}
\usepackage[table]{xcolor}
\usepackage[backend=biber,style=ieee,]{biblatex}
\usepackage{geometry}
\usepackage{fancyhdr}

\fancypagestyle{firststyle}{
  \fancyhf{}
  \fancyhead[C]{2026 Asia Pacific Signal and Information Processing Association Annual Summit and Conference (APSIPA ASC)}
}

\begin{document}

\title{Decoder-Side Semantic Conditioning for Low-Bitrate Neural Speech Compression}

\author{
\authorblockN{
Liuyang Bai, Weiyi Lu, Li Guo\authorrefmark{1}
}
\authorblockA{
Department of Computer Science and Data Science, New York University Shanghai, Shanghai, China
}
\authorrefmark{1}
Corresponding author
}

\maketitle
\thispagestyle{firststyle}
\pagestyle{empty}

\begin{abstract}
    Speech codecs are usually optimized for waveform fidelity, allocating bits to acoustic detail that can be inferred from linguistic structure. This leads to inefficient compression and degraded recognition performance. We propose SemDAC, a semantic-aware neural speech codec that adds hierarchical semantic conditioning to residual vector quantization (RVQ). The first RVQ quantizer is distilled from HuBERT features to produce semantic tokens capturing phonetic content, while later quantizers encode residual acoustics. The decoder is conditioned on semantic tokens via feature-wise linear modulation (FiLM), steering reconstruction toward information not explained by semantic abstraction. At 0.95 kbps, SemDAC matches or surpasses a 2.5 kbps DAC baseline on PESQ, STOI, SI-SNR, and Whisper WER, with comparable ViSQOL, and achieves higher subjective MOS than higher-bitrate DAC baselines. Results show that explicit semantic conditioning, rather than token disentanglement or increased model size alone, improves compression efficiency and recognition robustness.
\end{abstract}


\section{Introduction}

Audio and speech compression has long been a cornerstone of digital signal processing, driven by the need to reduce bandwidth while preserving perceptual quality. Traditional codecs—such as MP3~\cite{brandenburg1999mp3} and linear predictive coding~\cite{tremain1976linear}—rely on handcrafted features, parameter tuning, and extensive listening tests to achieve acceptable performance.  

Recently, neural audio codecs powered by deep learning have emerged as a powerful alternative \cite{zeghidour2021soundstream, defossez2022high}. These models adopt an encoder–quantizer–decoder architecture and learn compact audio representations directly from data. A key innovation is residual vector quantization (RVQ), which chains multiple vector quantizers to represent audio at progressively finer levels of detail \cite{zeghidour2021soundstream, defossez2022high, kumar2023high}. State-of-the-art codecs such as DAC~\cite{kumar2023high} achieve impressive fidelity at low bitrates, but their tokens are optimized for acoustic detail, leaving semantic information underrepresented. As a result, the decoder reconstructs audio solely from acoustically motivated latents, which can be inefficient for high-quality reconstruction and downstream tasks such as automatic speech recognition (ASR) and speech-language modeling \cite{borsos2023audiolm,rubenstein2023audiopalm}. 

Beyond DAC, several works have advanced codec design along different dimensions: HiFi-Codec~\cite{yang2023hifi} improves fidelity with group-RVQ, AudioDec~\cite{wu2023audiodec} targets low-latency streaming, MDCTCodec~\cite{jiang2024mdctcodec} combines MDCT with RVQ for lightweight coding, and LMCodec~\cite{jenrungrot2023lmcodec} employs causal transformers for ultra-low-bitrate speech. While effective, these advances continue to focus primarily on acoustic fidelity rather than semantic structure.  \looseness=-1

A key observation, overlooked by prior work, is that speech is inherently structured. Much of its acoustic variability is conditionally dependent on linguistic content. Phonemes largely determine spectral structure, while speaker-specific timbre and prosody provide variations around it. Self-supervised speech models such as HuBERT~\cite{hsu2021hubert} and wav2vec 2.0~\cite{baevski2020wav2vec} capture phonetic and semantic information at extremely low bitrates, proving highly effective for ASR \cite{borsos2023audiolm,rubenstein2023audiopalm}. This suggests that semantic representations could serve as strong priors for speech reconstruction, potentially reducing redundancy in acoustic coding. Recent semantic codec designs further support the usefulness of semantic representations for low-bitrate audio coding. For example, SpeechTokenizer~\cite{zhang2023speechtokenizer} disentangles semantic and acoustic tokens for speech representation learning, while SemantiCodec~\cite{liu2024semanticodec} introduces a separate semantic encoder to obtain compact semantic audio tokens at ultra-low bitrates. These works demonstrate that semantic information is valuable for compression and generation. However, they primarily treat semantic tokens as additional or disentangled representations.

In this work, we propose the \textit{Semantic Descript Audio Codec (SemDAC)}, a simple yet effective approach that introduces hierarchical semantic conditioning into neural speech compression. We designate the first layer of the residual vector quantization (RVQ) stack as a semantic quantizer, distilled from pretrained HuBERT embeddings, while subsequent layers encode residual acoustic details. Crucially, we condition the decoder on the semantic tokens using feature-wise linear modulation (FiLM)~\cite{perez2018film}. This decoder-side conditioning differs from prior semantic-token codecs: instead of using semantic tokens only as parallel features or disentangled code streams, SemDAC uses them as structural priors that guide the reconstruction process. The resulting architectural asymmetry encourages acoustic quantizers to represent residual variability relative to linguistic content, rather than redundantly encoding information already explained by semantic abstraction.

From a rate–distortion perspective, decomposing speech into semantic tokens $s$ and acoustic residuals $a$ naturally aligns with the entropy chain rule $H(s, a) = H(s) + H(a|s)$~\cite{cover1999elements}. By capturing high-level linguistic structure, semantic tokens provide prior information that reduces the conditional entropy of the acoustic residual $H(a|s)$, thereby enabling more efficient compression. 

\begin{figure*}[t] 
    \centering
    \includegraphics[width=0.8\linewidth]{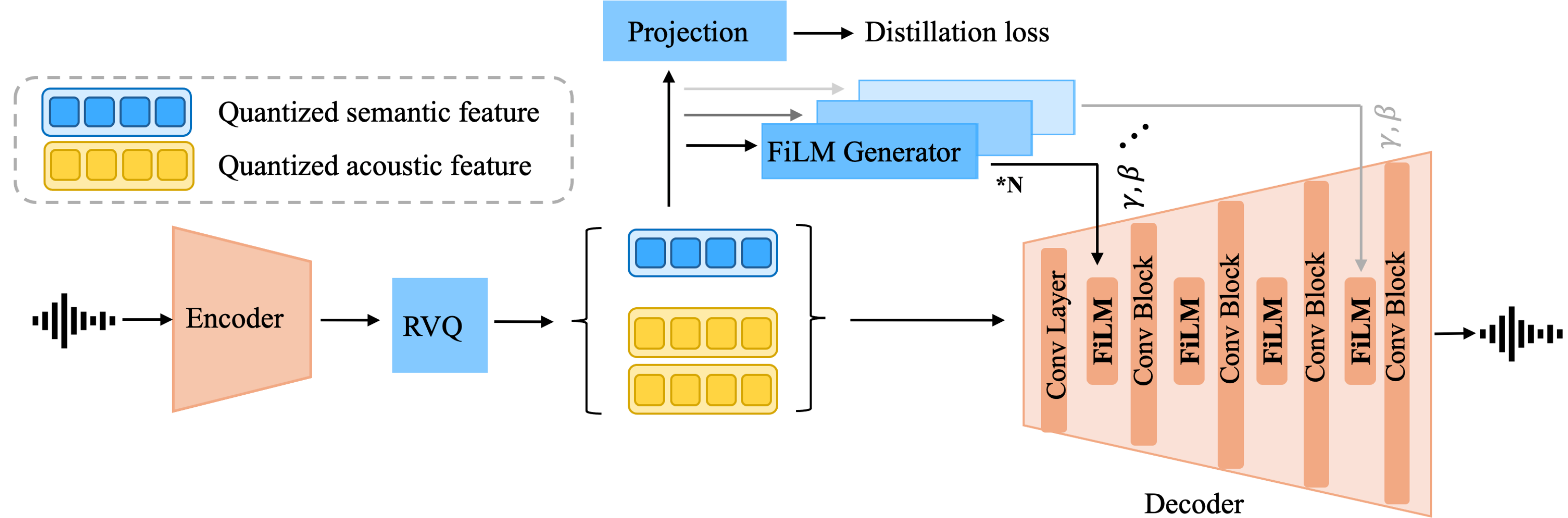}
    \caption{Architecture of SemDAC. The quantizer stack is divided into a semantic quantizer, supervised by a pretrained HuBERT model, and acoustic quantizers that encode residual details. FiLM generators map semantic tokens into modulation parameters, which are injected through FiLM modules into the decoder to enforce semantic consistency during reconstruction. \looseness=-1}

    \label{fig:model}
\end{figure*}

We demonstrate that this conditional design substantially improves compression efficiency. At extremely low bitrates (e.g., 0.95 kbps), 
SemDAC matches or surpasses a 2.5 kbps DAC baseline across most objective metrics, with comparable or higher MOS, while also achieving lower word error rate when evaluated with Whisper~\cite{radford2023robust} on reconstructed speech. Ablation studies further show that the primary gains arise from hierarchical conditional decoding rather than semantic token extraction. These findings indicate that explicit decoder conditioning, rather than token disentanglement alone, is the critical mechanism for leveraging semantic structure in neural speech compression.

\section{Methods}
\label{sec:methods}

Our proposed \textit{Semantic Descript Audio Codec (SemDAC)} builds upon the DAC framework~\cite{kumar2023high}, which follows the standard encoder–quantizer–decoder paradigm. The overall architecture is shown in Figure~\ref{fig:model}.
For a fair comparison, we retain the encoder and discriminator designs of DAC. SemDAC differs from the baseline model in two key ways: (i) it disentangles semantic from acoustic codes (\S\ref{sub:sem}), and (ii) it conditions the decoder on semantic priors (\S\ref{sub:film}). Importantly, our empirical results show that semantic distillation alone yields negligible improvements; substantial gains are observed only when semantic tokens are incorporated as an explicit conditioning signal in the decoder (Table~\ref{tab:film}). 
This asymmetric design introduces a hierarchical inductive bias that encourages acoustic representations to encode residual information relative to semantic structure, leading to improved compression efficiency and recognition performance. \looseness=-1

\subsection{Semantic Quantizer}
\label{sub:sem}

In DAC, all quantizers are treated uniformly, each modeling progressively finer acoustic residuals. In contrast, \textit{SemDAC} adopts an asymmetric design by designating the first quantizer as a semantic codebook, supervised by a pretrained HuBERT model~\cite{hsu2021hubert}. We use 9th-layer HuBERT features as the semantic teacher. The 9th layer is chosen as it provides a strong phonetic abstraction while retaining sufficient local acoustic cues.
A lightweight projection head maps semantic latents into the HuBERT feature space, and we minimize their Euclidean distance to the corresponding HuBERT feature embeddings at each time step. This \emph{semantic distillation loss} aligns the latent space with phonetic structure and transfers semantic knowledge into the codebook:
\begin{equation}
\mathcal{L}_{sem} = \frac{1}{T} \sum_{t=1}^{T} \big\lVert P(z^{\text{sem}}_t) - h_t \big\lVert_2^2 ,
\end{equation}
where $z^{\text{sem}}_t$ and $h_t$ denote the semantic latent and HuBERT embedding at time $t$, and $P(\cdot)$ is the projection layer.

The remaining RVQ layers serve as acoustic quantizers, modeling residual details not captured by the semantic tokens. Since semantic tokens mainly encode phonetic structure, they require fewer codewords. We therefore use 256–512 entries for the semantic codebook, while each acoustic quantizer employs 1024 entries. \looseness=-1

\subsection{FiLM-Conditioned Decoder}
\label{sub:film}

In DAC~\cite{kumar2023high}, the decoder is implemented as a symmetric counterpart to the encoder, reconstructing the waveform directly from the latent codes. This symmetric design is limited in efficiency, as it does not exploit the higher-level semantic information available in speech.

In SemDAC, we enhance the decoder by explicitly incorporating semantic guidance. 
The semantic and acoustic codes are first concatenated along the channel dimension and passed into the decoder, which consists of a pre-convolutional layer followed by four convolutional upsampling blocks for waveform reconstruction.

To further inject semantic information hierarchically, we introduce a FiLM generator implemented as a stack of convolutional layers that projects the semantic latents $z^{sem}_t$ to modulation parameters $\gamma_t$ and $\beta_t$. When necessary, these parameters are temporally upsampled to match the resolution of the decoder feature maps.
FiLM modulation can be flexibly inserted at selected stages of the decoder to scale and shift the intermediate combined feature representations. Specifically, given a decoder feature map $z_t$, the modulation is defined as
\begin{equation}
\text{FiLM}(z_t) = \gamma_t(z^{sem}) \odot z_t + \beta_t(z^{sem}),
\end{equation}
where $\odot$ denotes element-wise multiplication.

This design is flexible, allowing FiLM to be inserted at different locations within the decoder. Through extensive ablation studies, we find that placing the FiLM block between the pre-convolutional layer and the first upsampling block yields the best performance. Injecting semantic guidance at this early stage allows semantic information to shape the high-level structure of the audio and propagate through subsequent decoding stages. \looseness=-1

\subsection{Training Objective}
\label{sub:loss}

We adopt the same training objective as DAC~\cite{kumar2023high}, which combines multi-scale mel-spectrogram losses, adversarial and feature-matching losses from multi-period discriminators~\cite{kong2020hifi}, and standard codebook/commitment losses. To incorporate semantic guidance, we add a distillation loss that aligns the first quantizer’s latent codes with HuBERT embeddings (\S\ref{sub:sem}). \looseness=-1

The final loss is a weighted sum of all terms, with weights set to 15.0 for the multi-scale mel loss, 2.0 for the feature-matching loss, 1.0 for the adversarial loss, 1.0 and 0.25 for the codebook and commitment losses, respectively, and 1.0 for the semantic distillation loss.%

\begin{table*}[t]
\centering
\caption{Objective evaluation of the proposed codec at varying bitrates, compared with competing approaches.}
\label{tab:eval}
\begin{tabular}{lcc|ccccc}
\hline
\textbf{Model} & \textbf{Bitrate (kbps)} & \textbf{Parameters} & \textbf{PESQ $\uparrow$} & \textbf{STOI $\uparrow$} & \textbf{ViSQOL $\uparrow$} & \textbf{SI-SNR $\uparrow$} & \textbf{WER (\%) $\downarrow$} \\
\hline
Opus           & 3 & N/A  &   1.39  & 0.73 & 1.25 & -6.76 & 8.69   \\
Opus           & 6 & N/A  &   2.19  & 0.90 & 2.09 & 3.66 & 5.31   \\
\hline
DAC (retrain) & 1 & 73.9M   & 2.11 & 0.90 & 2.45 &  1.05 & 8.14 \\
DAC (retrain) & 2 & 74.0M   & 2.74 & 0.94 & 3.07 &  4.44 & 4.87 \\
DAC (retrain) & 2.5 & 74.0M   & 2.87 & 0.95 & 3.25 &  5.49 & 4.58 \\
DAC (retrain) & 3 & 74.0M   & 3.04 & 0.95 & 3.33 &  5.96 & 4.47 \\
\hline
SpeechTokenizer & 6 & 104M & 2.60 & 0.92 & 3.01 & 0.88 & 4.66 \\
SemantiCodec & 0.35 & 2.04B & 1.38 & 0.78 & 1.46 & -- & 35.97 \\
SemantiCodec & 0.7 & 1.03B & 1.77 & 0.85 & 2.13 & -- & 8.41 \\
SemantiCodec & 1.4 & 595M & 2.02 & 0.88 & 2.92 & -- & 5.84 \\
\hline
SemDAC (ours)  & 0.95 & 75.1M & 2.93 & 0.95 & 3.16 &  6.47 & 4.45 \\
SemDAC (ours)  & 1.95 & 75.2M & 3.24 & 0.96 & 3.52  &  7.08 & \textbf{4.20} \\
SemDAC (ours)  & 2.95 & 75.3M & \textbf{3.41} & \textbf{0.97} & \textbf{3.71}  &  \textbf{8.07} & 4.30 \\
\hline
\end{tabular}
\par\vspace{2pt}
{\footnotesize  SI-SNR is omitted for SemantiCodec because its diffusion decoder produces output that is not waveform-aligned with the reference, so sample-level SI-SNR reflects misalignment rather than quality.}
\end{table*}

\section{Experiment}
\label{sec:experiment}

\subsection{Experiment settings}

\textbf{Datasets}.  
To evaluate the effectiveness of semantic priors for speech compression, we train SemDAC on the LibriSpeech corpus~\cite{panayotov2015librispeech}, a widely used 360-hour benchmark for speech representation learning. During training, we extract 0.38-second excerpts from the audio and normalize them to \(-24\) dB LUFS to ensure consistent loudness.

\textbf{Model and training recipe}.  
We adopt DAC-16kHz~\cite{kumar2023high} as our baseline. Both the encoder and decoder consist of a pre-convolutional layer followed by four convolutional blocks, arranged symmetrically with downsampling rates [2, 4, 5, 8] and upsampling rates [8, 5, 4, 2], respectively.  
SemDAC retains this overall architecture but replaces the uniform quantization scheme with a semantic quantizer (codebook size 512) followed by acoustic quantizers (codebook size 1024). We also investigated FiLM conditioning at different decoder positions and found the best performance when inserting the FiLM block between the pre-convolutional layer and the first decoder block (\S\ref{sub:exp}). \looseness=-1  
 
Following DAC~\cite{kumar2023high}, we used a multi-period discriminator with periods [2, 3, 5, 7, 11]. For the reconstruction loss, we minimize the distance between log-mel spectrograms computed with window sizes [32, 64, 128, 256, 512, 1024, 2048], paired with corresponding mel bins [5, 10, 20, 40, 80, 160, 320]. The hop length is set to $1/4$ of the window length. In addition, we include feature-matching, codebook/commitment, and semantic distillation losses (\S\ref{sub:loss}). 
For semantic supervision, we use the HuBERT \texttt{base-ls960} checkpoint from Hugging Face, trained on the 960-hour LibriSpeech dataset. Specifically, we extract features from the 9th HuBERT layer as semantic targets.
All models were trained for 250k iterations on 0.38-second audio excerpts with a batch size of 48. These short excerpts are used only as efficient random training crops, following the DAC-style training setup, and do not constrain the duration of evaluation samples. All objective evaluation, ASR evaluation, and subjective listening tests are conducted on substantially longer utterances. Optimization is performed using AdamW~\cite{loshchilov2017decoupled} with a learning rate of $3\times10^{-4}$, $\beta_1=0.8$, and $\beta_2=0.9$.

\textbf{Evaluation Metrics}. 
We evaluate model performance using objective metrics commonly adopted in speech coding, including ViSQOL~\cite{chinen2020visqol}, PESQ~\cite{recommendation2001perceptual}, STOI~\cite{taal2010short}, and scale-invariant signal-to-noise ratio (SI-SNR). 
To assess downstream recognition robustness, we run Whisper (``medium.en'')~\cite{radford2023robust} on reconstructed speech and report word error rate (WER). 
In addition, we conduct a subjective listening study following a single-stimulus absolute category rating (ACR) protocol consistent with ITU-T P.800. We select 10 speech utterances, each approximately 15 seconds long, and process each utterance with each evaluated codec. The resulting 40 samples and 10 unprocessed references are randomized before being presented to 20 participants. Listeners rate each sample on a 1--5 scale, where 1 indicates completely unclear or severely distorted speech and 5 indicates completely clear and natural speech. Thus, all evaluated systems are judged on the same utterances under the same rating protocol.

\begin{figure}[t]
    \centering
    \includegraphics[width=0.95\linewidth]{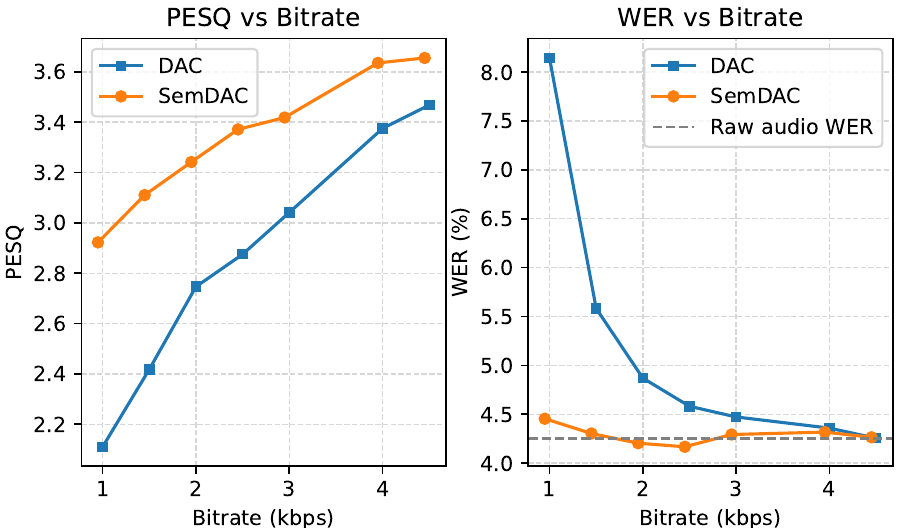}
    \caption{Bitrate–quality trade-off of SemDAC versus DAC, evaluated with PESQ and WER.}
    \label{fig:all}
\end{figure}

\begin{table}[t]
\centering
\caption{Subjective listening test (MOS) with 20 listeners on a 1--5 ACR scale (ITU-T P.800). \textit{Reference} denotes the original uncompressed audio, presented blind as an upper anchor. Values are mean MOS $\pm$ 95\% confidence intervals.}
\label{tab:mos}
\begin{tabular}{l c|c}
\hline
\textbf{Model} & \textbf{Bitrate (kbps)} & \textbf{MOS} $\uparrow$ \\
\hline
\rowcolor{gray!10} Reference & -- & $4.61\pm0.13$ \\
\hline
DAC (retrain)   & 1.00 & $3.86\pm0.43$ \\
DAC (retrain)   & 2.00 & $4.20\pm0.23$ \\
\hline
SemDAC (ours)   & 0.95 & $4.27\pm0.34$ \\
SemDAC (ours)   & 1.95 & $\mathbf{4.47\pm0.24}$ \\
\hline
\end{tabular}
\end{table}

\subsection{Experimental Results}
\label{sub:exp}

\textbf{Comparison to other methods}.
We compare \textit{SemDAC} with the traditional codec Opus~\cite{valin2012definition}, SpeechTokenizer~\cite{zhang2023speechtokenizer}, SemantiCodec~\cite{liu2024semanticodec}, and the neural codec DAC across a range of bitrates. Results are summarized in Table~\ref{tab:eval}. For fair comparison, ``DAC (retrain)'' denotes a DAC model trained from scratch under the same dataset, sampling rate, loss functions, optimizer settings, and RVQ configuration used for SemDAC, except for the proposed semantic quantizer and FiLM-based decoder conditioning; this isolates the effect of semantic conditioning from differences in data, training recipe, or implementation. SpeechTokenizer and SemantiCodec are evaluated from their official released checkpoints without retraining. We additionally report the number of model parameters for the neural codecs, allowing us to assess parameter efficiency alongside rate--distortion performance; since Opus is a hand-engineered codec rather than a learned model, its parameter count is marked N/A.

At matched bitrates, SemDAC consistently outperforms all baselines across objective metrics, with particularly strong gains in the low-bitrate regime where conventional codecs degrade most severely. Notably, SemDAC at 0.95~kbps matches or surpasses DAC at 2.5~kbps on PESQ, STOI, SI-SNR, and WER, with comparable ViSQOL, and is comparable to DAC at 3~kbps overall, demonstrating that SemDAC preserves both perceptual quality and intelligibility at substantially lower bitrates. SemDAC at 0.95~kbps also outperforms 6~kbps SpeechTokenizer~\cite{zhang2023speechtokenizer} on PESQ, STOI, SI-SNR, and WER despite its far lower bitrate. As SemantiCodec is a diffusion-based codec targeting general audio, we read this comparison primarily through recognition robustness and model size; under the reported metrics, SemDAC at 0.95~kbps achieves 2.93 PESQ, 0.95 STOI, and 4.45\% WER, while SemantiCodec at 1.4~kbps obtains 2.02 PESQ, 0.88 STOI, and 5.84\% WER. The parameter counts further highlight the efficiency of SemDAC. SemDAC contains approximately 75M parameters, only slightly larger than the retrained DAC baseline, while SpeechTokenizer contains 104M parameters and the evaluated SemantiCodec variants range from 595M to 2.04B parameters. Thus, the gains of SemDAC are not due to a substantially larger model, but to the way semantic tokens are integrated into the decoder.  \looseness=-1

Figure~\ref{fig:all} further illustrates the trends in PESQ and WER as a function of bitrate. SemDAC consistently outperforms DAC at matched bitrates across the full range. Once the bitrate exceeds 2~kbps, the WER of SemDAC matches that of raw audio (4.25\%) to within evaluation noise, indicating that reconstruction preserves intelligibility essentially fully. We attribute the small differences from the raw-audio reference at higher bitrates to run-to-run variance in Whisper decoding and finite-test-set sampling, and therefore interpret SemDAC's WER as saturating at the raw-audio level rather than exceeding it. The overall advantage of SemDAC over DAC remains clear throughout.

\textbf{Subjective quality.}
Objective metrics do not always fully capture perceived naturalness, especially at very low bitrates. We therefore also report MOS from a 20-listener subjective test (Table~\ref{tab:mos}), which includes the original uncompressed audio as a blind \textit{Reference} condition to anchor the rating scale. The MOS results are consistent with the objective trends in Table~\ref{tab:eval}: at 0.95~kbps SemDAC matches the DAC baseline operating at more than twice its bitrate, and at 1.95~kbps it approaches the unprocessed reference. This indicates that decoder-side semantic conditioning reduces perceptual artifacts while preserving intelligibility at substantially lower bitrates.

\textbf{Ablation Study.}  
We perform ablation experiments to evaluate the contribution of different design choices in SemDAC. Unless otherwise specified, all ablations use an RVQ with four quantization layers, consisting of one semantic quantizer and three acoustic quantizers.

\textit{Semantic codebook size.}  
We evaluate the effect of semantic codebook size by varying the number of entries in the first quantizer among 128, 256, 512, and 1024 (Table~\ref{tab:sem}). Both 256 and 512 entries strike a good balance between compactness and performance\footnote{Although 256 entries yield slightly better performance in the ablation study, we use 512 entries in main experiments for consistency across completed runs.}. Reducing the size to 128 substantially degrades quality, while increasing beyond 512 offers no further improvement.

\begin{table}[t]
\setlength{\tabcolsep}{3pt} 
\centering
\caption{Ablation study on the codebook size of the semantic quantizer.}
\label{tab:sem}
\begin{tabular}{c c|ccccc}
\hline
\textbf{Codebook size} & \rotatebox{90}{\shortstack{\textbf{Bitrate} \\ \textbf{(kbps)}}} & \rotatebox{90}{\textbf{PESQ}} & \rotatebox{90}{\textbf{STOI}} & \rotatebox{90}{\textbf{ViSQOL}} & \rotatebox{90}{\textbf{SI-SNR}} & \rotatebox{90}{\textbf{WER (\%)}} \\
\hline
1024      & 2.0    & 3.29   & 0.96 & 3.43 & 7.45 & 4.23 \\
\textbf{512 (default)} & 1.95 & 3.24   & 0.96 & 3.52 & 7.08 & 4.20 \\
256          &1.9   & 3.28   & 0.96 & 3.52 &  7.67 & 4.31 \\
128          &1.85  & 3.18   & 0.95 & 3.37 &  6.96 & 4.37 \\
\hline
\end{tabular}
\end{table}

\textit{FiLM conditioning.}  
We evaluate the role of FiLM conditioning by comparing models trained with and without FiLM layers, as well as variants where FiLM is inserted at different points in the decoder (Table~\ref{tab:film}). The semantic quantizer is fixed to the default codebook size of 512 in all cases. In the model without FiLM, the first quantizer is still distilled from HuBERT, and the semantic and acoustic codes are still concatenated before decoding. However, the decoder receives these tokens only as ordinary input features, rather than using the semantic stream as an explicit conditioning signal.

This distinction is central to the contribution of SemDAC. SemDAC without FiLM performs similarly to the DAC baseline, despite using the same semantic distillation objective. In contrast, adding FiLM conditioning at the earliest decoder location (F0) improves PESQ from 2.72 to 3.24 at approximately the same bitrate and also improves WER from 4.83\% to 4.20\%. These results show that the performance gain does not come merely from extracting semantic tokens or disentangling the first RVQ layer. Instead, the gain comes from using semantic tokens as decoder-side priors that modulate reconstruction.

For models trained with FiLM layers, semantic tokens are most effective when injected between the pre-convolutional layer and the first decoder block (SemDAC F0). Injecting semantics at this early stage provides a phonetic scaffold that shapes the high-level structure of the signal before subsequent upsampling stages synthesize finer acoustic details. By contrast, injecting semantics later in the decoder reduces their impact, since much of the acoustic structure has already been formed. Moreover, adding FiLM layers at multiple positions yields no consistent additional benefit, performing at best on par with a single early FiLM insertion.

\begin{table}[t]
\setlength{\tabcolsep}{3pt} 
\centering
\caption{Ablation study on FiLM placement. 
F0 indicates that FiLM is inserted between the pre-convolutional layer and the first decoder block. 
F$i$ denotes FiLM insertion before the $i$-th decoder block. 
``+'' indicates multiple FiLM insertions.}
\label{tab:film}

\begin{tabular}{l c|ccccc}
\hline
\textbf{Model} & \rotatebox{90}{\shortstack{\textbf{Bitrate} \\ \textbf{(kbps)}}} & \rotatebox{90}{\textbf{PESQ}} & \rotatebox{90}{\textbf{STOI}} & \rotatebox{90}{\textbf{ViSQOL}} & \rotatebox{90}{\textbf{SI-SNR}} & \rotatebox{90}{\textbf{WER (\%)}} \\
\hline
\rowcolor{gray!10} DAC (baseline)     & 2.00 & 2.74 & 0.94 & 3.07 & 4.44 & 4.87 \\
\rowcolor{gray!10} SemDAC w/o FiLM    & 1.95 & 2.72 & 0.94 & 2.96 & 3.96 & 4.83\\ 
\rowcolor{gray!10} \textbf{SemDAC F0 (default)}& 1.95 & {3.24} & 0.96 & {3.52} & {7.08} & {4.20} \\
SemDAC F1          & 1.95 & 2.96 & 0.95 & 3.19  & 5.49 & 4.46 \\
SemDAC F2          & 1.95 & 2.83 & 0.94 & 3.18 & 4.38 & 4.68 \\
SemDAC F3          & 1.95 & 2.74 & 0.94 & 2.95   & 3.62 & 4.78 \\
SemDAC F0+F1       & 1.95 & 3.22 & 0.96 & 3.37   & 7.14 & 4.39   \\
SemDAC F0+F2       & 1.95 & 3.22 & 0.96 & 3.43   & 6.94 & 4.35   \\

SemDAC F0+F3       & 1.95 & 3.36 & 0.96 & 3.42   & 7.49 & 4.20   \\
\hline
\end{tabular}
\end{table}

\textit{HuBERT layer choice.}
We ablate the choice of HuBERT layer used as the semantic teacher by extracting features from the 6th, 9th, and 12th layers under the default configuration (Table~\ref{tab:hubert_layer}). The 9th and 12th layers yield comparable performance, while the 6th layer is slightly worse, suggesting that mid-to-late HuBERT representations provide a better semantic prior for reconstruction.
\begin{table}[t]
\setlength{\tabcolsep}{4pt}
\centering
\caption{Ablation on the HuBERT layer used for semantic distillation (in the default SemDAC setting).}
\label{tab:hubert_layer}
\begin{tabular}{c|ccc}
\hline
\textbf{HuBERT layer} & \textbf{PESQ} $\uparrow$ & \textbf{SI-SNR} $\uparrow$ & \textbf{WER (\%)} $\downarrow$ \\
\hline
6  & 3.10 & 6.95 & 4.40 \\
\textbf{9 (default)}  & 3.20 & 7.10 & 4.20 \\
12 & 3.20 & 7.08 & 4.10 \\
\hline
\end{tabular}
\end{table}

\subsection{Discussion}

The ablation results indicate that the key contribution of SemDAC is not simply the introduction of semantic tokens, but their use as an explicit decoder-side conditioning signal. Semantic distillation alone produces a first RVQ layer that is more aligned with phonetic structure, but this alignment does not substantially improve reconstruction unless the decoder is architecturally encouraged to use it. FiLM provides this mechanism by allowing semantic tokens to scale and shift intermediate decoder features, thereby turning linguistic information into an active reconstruction prior.

This finding clarifies the relationship between SemDAC and prior semantic-token codecs. SpeechTokenizer~\cite{zhang2023speechtokenizer}, SemantiCodec~\cite{liu2024semanticodec}, and XCodec~\cite{ye2025codec} demonstrate that semantic representations are useful for low-bitrate coding and audio-language modeling. SemDAC is complementary to these works: rather than asking only whether semantic tokens can be extracted, it studies how they should be integrated into the codec decoder. The FiLM ablation in Table~\ref{tab:film} shows that treating semantic tokens as ordinary concatenated inputs is insufficient, while using them as hierarchical conditioning signals leads to substantially better rate--distortion behavior.

Conditioning the decoder on semantic tokens provides explicit phonetic scaffolding for waveform reconstruction, allowing the acoustic quantizers to focus on fine-grained details such as timbre and prosody. This disentanglement and
collaboration between semantic and acoustic representations yields more accurate and efficient reconstructions, improving both perceptual quality and intelligibility at lower bitrates. \looseness=-1

\section{Conclusion}
In this paper, we demonstrated that incorporating semantic priors provides a powerful inductive bias for neural speech codecs. By guiding the decoder with semantic tokens distilled from a pretrained HuBERT model, SemDAC achieves more efficient use of bits, yielding both higher perceptual quality and improved recognition accuracy. Importantly, the gains arise not simply from distilling semantic features, but from explicitly integrating them into the decoding process, underscoring the value of semantic guidance in neural audio compression. \looseness=-1
\newpage

\printbibliography

\end{document}